\documentclass[pra,aps,showpacs,epsfig,amsmath,amsthm,amsfonts,amssymb,twocolumn]{revtex4}

\usepackage{graphicx}

\begin{document}
\title{Quantum Interference in the Kirkwood-Rihaczek representation}
\author{L. Praxmeyer and K. W\'odkiewicz}
\affiliation{Institute of Theoretical Physics,
Warsaw University,\\ 
Ho\.za 69, 00--681 Warsaw, Poland }

\date{10 July 2002 }

\begin{abstract}
We discuss the Kirkwood-Rihaczek phase space distribution
and analyze a whole new class of quasi-distributions  connected with this
function. All these functions have the correct marginals.
We construct a coherent state representation of such functions, discuss
which operator ordering corresponds to the Kirkwood-Rihaczek distribution
and their generalizations, and show how such states are connected to squeezed
states. Quantum interference in the Kirkwood-Rihaczek representation
is discussed.
\end{abstract}


\maketitle

\section{Introduction}
A number of different phase space distribution functions has been
introduced and investigated over the years, the Wigner
distribution function \cite{Wigner} being the most famous and the
most known of all. The phase space distributions from the
Glauber-Cahill s-parameterized class of quasi-distributions
\cite{GlaCah} that contains the Wigner function, the
 Glauber-Sudarshan $P$-representation \cite{Gla,Sud}, and the
Husimi or the $Q$-representation \cite{Husimi}, have been widely used
as useful and powerful  phase space tools.

It is the purpose of this paper to show that a different class of phase space
 distribution functions,
connected with a lesser-known Kirkwood distribution function
\cite{Kirkwood}, can be useful in the investigation of quantum
interference in phase space.  The Kirkwood
distribution was proposed just a year after Wigner introduced his
function and like the Wigner distribution  was firstly used in
quantum statistics and thermodynamics. The Kirkwood distribution
has been rediscovered by Rihaczek \cite{Rihaczek} for use in the
theory of time-frequency analysis of classical signals. Operator
bases of Kirkwood and Wigner distribution functions have been
examined by Englert \cite{Englert}. Zak \cite{Zak} studied the
Rihaczek function in the quasimomentum-quasicoordinate
representation.

In this paper we shall study a new class of phase space distribution functions
connected with the Kirkwood-Rihaczek distribution function.
 The main advantage of the functions from this class is that they all lead to
the correct position and momentum marginals. We shall show how to
construct a coherent state representation of such functions and which
operator ordering corresponds to the Kirkwood-Rihaczek
distribution and their generalizations.
A connection of such phase space
functions with squeezed states will be established.

In Section \ref{sec2}  we present the basic definitions, properties, and
differences
of the Wigner and Kirkwood distributions. Section \ref{sec3} is devoted to the problem of operator ordering and its relation to various phase space
distributions. In Section \ref{sec4} the definition of
generalized Kirkwood-Rihaczek distribution function is introduced and it is
shown that Kirkwood-Rihaczek function  fully characterize the quantum state.
This means that
a full quantum state reconstruction with the help of the Kirkwood-Rihaczek function is possible.
 Sections \ref{sec5} and \ref{sec6} are devoted
to the quantum interference in phase space. Using simple examples, we present
various properties of the Kirkwood-Rihaczek functions and we discuss similarities and differences of such phase space quasi-distributions with the Wigner function.

\section{Phase space and quantum marginals}\label{sec2}
\subsection{The Wigner distribution function}
In 1932 Wigner introduced a phase-space distribution function
\begin{equation}
W_{\Psi}(q,p)=
\frac{1}{2\pi\hbar}\;\int
\Psi^{\star}(q+\xi/2)\; e^{\frac{ip\xi}{\hbar}}\;\Psi(q-\xi/2) d\xi\;,
\label{zwig}
\end{equation}
which fulfills the fundamental requirement for a joint probability
distribution  in phase space, i.e.  when integrated over $q$ or $p$, it
 gives marginal probabilities: $ |\Psi(q)|^{2} $ and $\frac{1}{2\pi\hbar}\;
|\widetilde{\Psi}(p)|^{2} $,
where
$
\widetilde{\Psi}(p)=\int dq\, e^{-\frac{ipq}{\hbar}}\,\Psi(q)\;
$
is the Fourier transform of $\Psi(q)$.

As shown by Wigner, the function given by Eq. (\ref{zwig}) is in
general not positive. However, under simple and reasonable
physical
 assumptions it is unique. Thus, when one demands that:
(i) a phase space distribution $P(q,p)$ is real, (ii) bilinear in $\Psi$, (iii)
gives the correct marginals, and (iv) its dynamical evolution     reproduces
the Liouville equation in the classical limit, then the distribution $P(q,p)$
is unique and is the Wigner function \cite{O'Connell}.

After Wigner's work many different phase space
distribution functions have been introduced and investigated. There is a very rich
literature devoted to applications of the Wigner function and other  various
quasi-distributions in quantum optics (see e.g. \cite{ff}).

The simplest example of a phase space distribution which does not satisfy the
four assumptions  leading to the Wigner function, but reproduces the right
marginal in an explicit way, is the distribution function:
\begin{equation}
P(q,p)=\frac{1}{2\pi\hbar} |\Psi(q)|^{2}\;
|\widetilde{\Psi}(p) |^{2} .
\label{prop}
\end{equation}
Clearly, this distribution is not bilinear in  $\Psi$ and as a result does not
satisfy the requirements of the Wigner's uniqueness theorem. Another
distribution function similar to that given by Eq. (\ref{prop}) can be
 guessed taking formally a
square-root of this expression. As a result, up to an arbitrary phase, we have
\begin{equation}
P(q,p)\sim |\Psi(q)|\;
|\widetilde{\Psi}(p) | .
\label{prop1}
\end{equation}
Note that this distribution function contains no information about
the phase of the wave function.  A simple insertion  of  an
additional phase factor $\varphi(q,p)$ to the wave functions leads
to the expression:
\begin{equation}
P(q,p)\sim \Psi(q) \; e^{i\varphi(q,p)}\; \widetilde{\Psi}^{\star}(p)\,,
\label{prop2}
\end{equation}
that defines  a class of   bilinear but complex distribution
functions.  An example of such a distribution  has been proposed
just one year after Wigner introduced his famous  distribution
function.

\subsection{The Kirkwood-Rihaczek distribution function}
In 1933 Kirkwood  introduced
a phase space distribution which, according to his description, ``{\em \ldots
differs but little from the Wigner function}''\cite{Kirkwood}. The Kirkwood function is defined
as follows:
\begin{eqnarray}
K(q,p)&=& \frac{1}{2\pi\hbar} \int d\xi\, \Psi(q)\,
e^{\frac{i(\xi-q)p}{\hbar}}\,\Psi^{\star}(\xi)=\nonumber\\
& =&\frac{1}{2\pi\hbar}\, \Psi(q)\, e^{- \frac{ipq}{\hbar}}\,
\widetilde{\Psi}^{\star}(p).
\label{zrih}
\end{eqnarray}
In 1968, this function was rediscovered by Rihaczek in the context of a signal
energy distribution in time and frequency (for a review of time-frequency
distributions see \cite{Cohen0,Coh01}).

The real part of the Kirkwood-Rihaczek distribution
\begin{equation}
K_{{\mathrm{Re}}}(q,p)={\mathrm{Re}}\left[ \Psi(q)\, e^{- \frac{ipq}{\hbar}}\,
\widetilde{\Psi}^{\star}(p)\right] \label{re-zrih}
\end{equation}
 is closely related to a quantum mechanical phase space distribution
introduced by Margenau and Hill \cite{M-H}.

We clearly see that the Kirkwood-Rihaczek (K--R)  distribution has the correct
marginal properties:
\begin{equation}
 \int  K(q,p)\,dp\, =\, |\Psi(q)|^{2}\;,\nonumber
\end{equation}
\begin{equation}
 \int   K(q,p)\,dq\, =\,\frac{1}{2\pi\hbar}\;
|\widetilde{\Psi}(p)|^{2}\;,\label{kweew}
\end{equation}
and it is normalized,
\begin{equation}
 \iint K(q,p)\, dq\,dp=1,
\end{equation}
as a consequence of the normalization of the wave  function
$\Psi(q)$. Note that the contribution of the imaginary part of the
K--R distribution to the marginals is identically zero. Hence, we
shall often focus on the real part.

Another condition easy to find  is  that the  absolute square of
$K(q,p)$ has the form similar to the Eq.(\ref{prop}), i.e.
\begin{equation}
  \mid K(q,p)\mid^2\,=  \frac{1}{(2\pi\hbar)^2}\, |\Psi(q)|^{2}\;
|\widetilde{\Psi}(p) |^{2},\label{absq}
\end{equation}
which indicates that $
K(q,p)$ is a
 square-integrable function, with
\begin{equation}
 \iint \mid K(q,p)\mid^2  dq\,dp=\frac{1}{(2\pi\hbar)}.
\end{equation}
The absolute square of the K--R function, Eq. (\ref{absq}), has a simple
physical interpretation. It is just proportional to  the product of the
probabilities in configuration and momentum representations.

The  dynamical free evolution of a particle with mass $m$ of the K--R
distribution function is given by the following equation
\begin{equation}
\partial_{t} K(q,p,t)\;+\;
\frac{p}{m}\;\partial_{q}
K(q,p,t)=\frac{i\hbar}{2m} \partial_{q}^2
 K(q,p,t)\; ,
\label{ewol:rih} \end{equation}
which can be also written in a form:
\begin{equation*}
K(q,p,t)=e^{t(
\frac{i\hbar}{2m}\partial_{q}^2-
\frac{p}{m}\partial_{q})}K(q,p,0).
\end{equation*}
We see from this formula, that the free evolution of the K--R
function is a superposition of the free Schr\"odinger diffusion
and of a classical boost to a moving frame. The K--R distribution
function is bilinear and has the correct marginal properties but,
in contrast with the Wigner function, it is not real nor does its
free evolution satisfy the classical Liouville equation. As we
shall see in detail later, it is also not well-behaved under
rotations of the $(p,q)$ coordinate system, and this implies that
such a function  can not be measured by tomography methods.

The simplicity of the definition, Eq. (\ref{zrih}), indicates that it is
relatively easy to evaluate the K--R distribution function even
for systems for which an analytical formula of the Wigner function
is not known. The best example of such a system is a hydrogen
atom. Elsewhere  we shall present the K--R functions for different
energy levels of the  hydrogen atom \cite{new}.
\subsection{The Cohen  distribution functions}
We have already emphasized the marginal properties of a phase space
distribution
function.
An interesting question arises about a general form of the distribution with
the correct marginals. This problem  has been posed and solved by Cohen in 1966
\cite{Cohen,Cohen2}. The most general distribution $P(q,p)$ with the proper
marginals has  the form of a double Fourier transform of a function
\begin{equation}
A(q',p') = \int e^{-ip'\xi/\hbar}\;
\Psi^{\star}(\xi-q'/{2})\;\Psi(\xi+q'/{2})\;d\xi\; \label{zamb}
\end{equation}
multiplied by  an arbitrary  function $\Phi(q',p')$ satisfying the
relations
\begin{equation}
\Phi(q',0)=1=\Phi(0,p')\;.
\label{war}
\end{equation}
In the literature devoted to optical processing of classical signals, the
function $A(q',p')$  is called the Ambiguity function.

Therefore,  the Cohen joint distribution functions labeled by
functions
 $\Phi$ are given by the following equation
\begin{eqnarray}
&  & P_{\Phi}(q,p)={\cal F}\, [\,\Phi\, A\,]:=\label{uog:wig}\\
& &=\frac{1}{(2\pi\hbar)^{2}} \iint e^{i(p'q-q'p)/\hbar}
\Phi(q',p')A(q',p')dp'dq',\nonumber
\end{eqnarray}
where by ${\cal F}$ we have denoted a double Fourier transform.

The Wigner distribution function is obtained by substituting  $\Phi(q',p')=1$
 in the Cohen
distribution functions formula.  The K--R distribution function is obtained for
$\Phi(q',p')=\exp[-ip'q'/(2\hbar)]$;
 $\Phi(q',p')=\cos[p'q'/(2\hbar)]$ leads to the Margenau-Hill distribution.

\section{Wigner-Weyl transformation}\label{sec3}
\subsection{The $(q,p)$ phase space description}
The Wigner-Weyl association of classical phase space
 functions ${\cal A}(q,p)$ with quantum
operators follows form the property:
\begin{eqnarray}
 & &\{ {\cal A}(\hat{q},\hat{p})\}_{\rm ordering}=\\
 & & =\iint dq  dp\,
{\cal A}(q,p)\,\{ \delta(q-\hat{q})\delta(p-\hat{p}) \}_{\rm ordering}.
\nonumber
\end{eqnarray}
 The problem of defining a quantum operator in phase space lies in the ordering of operators  $\hat{q} $ and $
\hat{p}$. As an example of such association we shall take a phase space
density distribution, whose quantum average leads to the density operator of
the system.
 Using the Fourier decomposition of the Dirac delta functions we see
that  the ordering becomes equivalent to  an ordering of the Heisenberg-Weyl
algebra operators:
\begin{eqnarray}
& &\{\delta(q-\hat{q})\delta(p-\hat{p}) \}_{\rm ordering} =\label{dirac}\\
& &=\frac{1}{(2\pi\hbar)^2}
        \iint dq' dp'
        e^{\frac{i(p'q-q'p)}{\hbar}}
        \left\{ e^{\frac{i\hat{p}q'}{\hbar} }
        e^{-\frac{ip'\hat{q}}{\hbar} } \right\}_{\rm ordering}.\nonumber
\end{eqnarray}
This formula shows that there is no natural unique  generalization of the
classical probability density in phase space, because it is possible to have
arbitrary classes of operator orderings. It has been recognized by Wigner and
Weyl that various quantum distribution functions can be associated with
different orderings of operators $\hat{q}$ and $\hat{p}$.
 For example,
the Wigner distribution function corresponds to Wigner-Weyl ordering, which is
obtained by putting $\hat{q} $ and $ \hat{p}$ operators in the same exponent in
Eq. (\ref{dirac}). We shall show that the K--R distribution function
corresponds to a special ordering called the anti-standard ordering.

One can easily find that the Ambiguity function, Eq.(\ref{zamb}), may be expressed also as
\begin{equation}
A(q',p')= \langle \Psi| e^{\frac{i\hat{p}q'-ip'\hat{q}}{\hbar}}
|\Psi\rangle\;.
\label{a1}
\end{equation}
Using the above equation, the definition of the K--R distribution function can
be formulated as follows
\begin{equation}
K(q,p)
=\frac{1}{(2\pi\hbar)^{2}}
        \iint dp'dq'  e^{\frac{i(p'q-q'p)}{\hbar}}
        \langle \Psi| e^{\frac{i\hat{p}q'}{\hbar}}
e^{\frac{-ip'\hat{q}}{\hbar}}
        |\Psi\rangle\;.
\nonumber
\end{equation}
By comparison with Eq. (\ref{dirac}) we find that the Kirkwood distribution
function corresponds to an ordering  such that all $\hat{p} $ operators are on the left of all $\hat{q}$ operators. Such
an arrangement of the canonical operators is called the anti-standard ordering.
Analogously one can find that the complex conjugation of the K--R function
corresponds to the standard ordering (all operators $ \hat{q}$ are on the left
followed by  all $ \hat{p}$ operators). The real part of K--R distribution
function is associated with a symmetric superposition of the  anti-standard and
the standard ordering, i.e.
$\frac{1}{2}(\hat{q}^{n}\hat{p}^{m}+\hat{p}^{m}\hat{q}^{n})$. Note that such
an ordering is not equivalent to the Wigner-Weyl ordering leading to the Wigner
function.

\subsection{Coherent state phase space description}
So far various properties and definitions of phase space
distribution
 functions were formulated in position or momentum representations.
Such a parameterization seems to be the most natural  to study phase space
properties  of particles. However, Glauber and Cahill have pointed out that the
coherent state representation is more natural and useful while dealing with
phase space distribution functions describing the quantum states of light
\cite{GlaCah}.

We shall now use the following notation:
  Greek letters  ($\alpha,\; \beta$, etc.) designate complex variables;
 $\hat{a}$, $\hat{a}^{\dagger}$  annihilation
and creation operators;
 $\mathrm{\hat{D}}(\alpha)=
 \exp(\alpha \hat{a}^{\dagger} -\alpha^{\star} \hat{a})$
denotes displacement operator; and all integrations are taken over
the whole complex plane. Using the formalism of coherent states
Glauber and Cahill have shown that it is possible to define an
$s$-parameterized class of distribution functions simply related
to the Wigner distribution function. These quasi-distributions are
defined as a complex Fourier transform ${\cal F}$
 of the $s$-ordered characteristic function:
\begin{equation}
 W(\alpha,s)={\cal F}\,[\,C\,]  :=\int \frac{d^{2}\beta}{\pi^{2}}\;
e^{\alpha\beta^\star-\alpha^\star\beta} C(\beta,s),
\label{wigs}
\end{equation}
defined as
\begin{equation}
C(\beta,s)=e^{\frac{s-1}{2} |\beta|^{2}}
        \mathrm{Tr}\left[\hat{\rho}
        e^{\beta a^{\dagger}}e^{-\beta^{\star} a}\right]=
        e^{\frac{s}{2} |\beta|^{2}}
        \mathrm{Tr}[\hat{\rho}\mathrm{\hat{D}}(\beta)].
\end{equation}
In the equation above $\hat{\rho}$ is the  density operator of the investigated system.  The  normally ordered form of
$\mathrm{\hat{D}}(\beta)$ allows to perform the integral explicitly, and then Eq.
(\ref{wigs}) can be rewritten as
\begin{equation}
W(\alpha,s) = \frac{2}{\pi(1-s)} \mathrm{Tr}
\left[ \hat{\rho} \mathrm{\hat{D}}(\alpha)
\hat{\Pi}(s)
 \mathrm{\hat{D}}^{\dagger}(\alpha)\right]
\label{wig:hatn}
\end{equation}
with the operator $\hat{\Pi}(s)$ defined as:
\begin{equation}
\hat{\Pi}(s)=\left( \frac{s+1}{s-1} \right)^{\hat{a}^{\dagger}\hat{a}}.
\end{equation}
The continuous parameter $s$ (which is required to be real and to satisfy an
inequality $s \leq 1$) corresponds to differing ordering of the creation and
annihilation operators. Three values: $s=1$, $s=0$, and $s=-1$ correspond to
normal, symmetric, and anti-normal ordering, that lead to the Glauber-Sudarshan
$P$-representation,
the Wigner function, and the $Q$-representation,  respectively.

These results have been generalized by  Agarwal and Wolf in \cite{AgW}.
They have
proposed the following general formula for the quasi-distribution functions:
\begin{equation}
W(\alpha,\Omega)=\int\frac{d^{2}\beta}{\pi^{2}}
        e^{\alpha\beta^\star-\alpha^\star\beta}
        \Omega(\beta,\beta^{\star})
         \mathrm{Tr}\left[\hat{\rho} \mathrm{\hat{D}}(\beta)\right] \;,
\label{Omeg}
\end{equation}
where
$\Omega(\beta,\beta^{\star})$ is an analytic function  of the complex variable
 $\beta$ that has no zeros.
 The condition $\Omega(0,0)=1 $ guarantees the  normalization
\begin{equation}
\int d^2\alpha\, W(\alpha,\Omega)=1.
\end{equation}
Obviously, such functions  do not provide automatically  the correct quantum
marginals, and only for  very selected  functions from the class given by Eq.
(\ref{Omeg}), we can  satisfy the relations from Eq. (\ref{kweew}).

In the following Section we shall show that the K--R distribution
corresponds to a specific form of the $\Omega$ function. Moreover,
with the help of this formalism we shall define a new class of
$\sigma$-ordered K--R distribution functions in full analogy to
the $s$-ordered functions of Cahill and Glauber.

\section{Generalized K--R function and quadratic ordering}\label{sec4}
\subsection{Definition of generalized K--R distribution}
A new class of quasi-distributions involving a quadratic ordering function:
$\Omega(\beta,\beta^{\star})=\exp\left(\sigma(\beta^{\star\;2}-
\beta^{2})/4 \right) $
is defined as a Fourier transform
\begin{equation}
 K(\alpha,\sigma)= {\cal F}[C(\beta,\sigma)]=\int \frac{d^{2}\beta}{\pi^{2}}
e^{\alpha\beta^\star-\alpha^\star\beta} C(\beta,\sigma), \label{k2}
\end{equation}
of the characteristic function  given by
\begin{equation}
C(\beta,\sigma)=e^{\sigma\frac{\beta^{\star\;2}-\beta^{2}}{4}}
\mathrm{Tr}(\hat{\rho} \mathrm{\hat{D}}(\beta)) \,.
\label{RIH1}
\end{equation}
The continuous parameter $\sigma$ is real and there are no limits to its value.
The K--R distribution function corresponds to $\sigma=1$, for $\sigma=0$ we
again obtain the Wigner distribution function.

As in the case of s-ordered Cahill-Glauber
distribution functions, we can perform the integrals and then rewrite
Eq. (\ref{k2}) as
\begin{equation}
 K(\alpha,\sigma) =\frac{2}{\pi\sqrt{1+{\sigma}^2}}
\mathrm{Tr \left[ \hat{\rho}\;\hat{D}(\alpha)
 \hat{K}(\sigma)
\hat{D}^{\dagger}(\alpha) \right]},  \label{kkir}
\end{equation}
where
\begin{equation}
\mathrm{   \hat{K}(\sigma)    =\,:\,e^{\frac{-2\hat{a}\hat{a}^{\dagger}+
\sigma(\hat{a}^{\dagger}\hat{a}^{\dagger}-\hat{a}\hat{a})}{1+{\sigma}^2}}}\,:\,.
\label{oper}
\end{equation}
In the above formula $\;:\;\;:\;$ denotes normal ordering of operators
$\hat{a}$, $\hat{a}^{\dagger}$. Using the identity $t^{\hat{a}^{\dagger}\hat{a}}=
\,:e^{(t-1  )\hat{a}^{\dagger}\hat{a}}:\, $ we can transform Eq. (\ref{oper}) into
\begin{equation}
 \mathrm{} \hat{K}(\sigma) =
e^{\frac{\sigma\hat{a}^{\dagger}\hat{a}^{\dagger}}{1+{\sigma}^2}}
\left(\frac{{\sigma}^2 - 1}{1+{\sigma}^2} \right)^{\hat{a}^{\dagger} \hat{a}}
e^{-\frac{\sigma\hat{a}\hat{a}}{1+{\sigma}^2}}\;.\label{opkir}
\end{equation}

Equations (\ref{kkir}--\ref{oper}) define a $\sigma$-ordered class
of generalized K--R phase space distributions. From the property
$K^{\dagger}(\alpha,\sigma)=K(\alpha,-\sigma)$, it is clear that
the only real function in this class is the Wigner function which
corresponds to $\sigma=0$.

The operator $\mathrm{\hat{K}}(\sigma)  $ is not Hermitian but has a finite trace
\begin{equation}
\mathrm{Tr} \left[ \mathrm{ \hat{K}}(\sigma)\right]=\frac{\pi}{2}\sqrt{1+{\sigma}^2}.
\end{equation}
Let us define operators ${\cal\hat{K}}(\alpha,\sigma)$ as
\begin{equation}
{\cal\hat{K}}(\alpha,\sigma)=\mathrm{\hat{D}(\alpha)
 \hat{K}(\sigma)
\hat{D}}^{\dagger}(\alpha),
\end{equation}
and study its properties.
We find that sets ${\cal\hat{K}}(\alpha,1)$ and
 ${\cal\hat{K}}^{\dagger}(\alpha,1)$ form a
complete, orthogonal basis systems with respect to the scalar product
\begin{equation}
\mathrm{Tr}\left[{\cal\hat{K}}^{\dagger}(\alpha,1)\,{\cal\hat{K}}(\beta,1)\right]=\frac{\pi}{2}\,
\delta(\alpha-\beta).
\end{equation}
That means that the K--R phase space distribution, up to a normalization
factor, is an expansion coefficient of density
 operator in the basis
 ${\cal\hat{K}}^{\dagger}(\alpha,1)$.
An arbitrary operator $\hat{\rho}$ may be expanded as
\begin{equation}
\hat{\rho}=\sqrt{2}\int d^{2}\alpha
K(\alpha,1){\cal\hat{K}}^{\dagger}(\alpha,1),\label{roz}
\end{equation}
or, equivalently,
\begin{equation}
\hat{\rho}
=\sqrt{2}\int d^{2}\alpha K(\alpha,-1)\,
{\cal \hat{K}}(\alpha,1).
\end{equation}
From Eq. (\ref{roz}) is clear that knowledge of K--R distribution of the state  $\hat{\rho}$
is equivalent to the knowledge of the state itself. The same holds for the complex
conjugate of K--R distribution function.

\subsection{Connection with squeezed states}
From the formula given by Eq. (\ref{opkir}) we see that the K--R function
 corresponds to a special case when the
operator $\left(\frac{{\sigma}^2 - 1}{1+{\sigma}^2} \right)^{\hat{a}^{\dagger} \hat{a}} $
becomes a projector, as:
\begin{equation}
\lim _{\sigma\rightarrow 1} \left(\frac{{\sigma}^2 - 1}{1+{\sigma}^2}
\right)^{\hat{a}^{\dagger} \hat{a}}=| 0\rangle\langle 0|. \end{equation} Thus,
as regards the K--R distribution, the  formula given by Eq. (\ref{kkir}) is reduced
to
\begin{equation}
K(\alpha,1)=\frac{\sqrt{2}}{\pi}
\mathrm{Tr}\left[\hat{\rho}\;\mathrm{\hat{D}}(\alpha)
e^{\frac{\hat{a}^{\dagger}\hat{a}^{\dagger}}{2}}
| 0\rangle\langle 0| e^{-\frac{\hat{a}\hat{a}}{2}}
\mathrm{\hat{D}}^{\dagger}(\alpha)\right]  .
\label{rih:0}
\end{equation}
The above equation may be rewritten in more intuitive form if we refer to
the definition of the  squeezed state \cite{squeeze}. The general coherent squeezed state
$|\alpha,\xi\rangle=\mathrm{\hat{D}}(\alpha)\hat{S}(\xi)|0\rangle$ is obtained by a
combined action of the displacement and the squeezing operator
\begin{equation}
 \hat{S}(\xi) =\exp\left(-\frac{\xi}{2}\hat{a}^{\dagger 2}+
\frac{\xi^{\star}}{2}\hat{a}^{2}\right)
\end{equation}
 on the vacuum state.
The squeezing operator can be written in an ordered form
\begin{equation}
\hat{S}(\xi) =\exp\left(-\frac{\nu}{2\mu}\hat{a}^{\dagger 2}\right)
\left(\frac{1}{\mu}\right)^{\hat{a}^{\dagger}\hat{a}+\frac{1}{2}}
\exp\left(\frac{\nu}{2\mu}\hat{a}^{2}\right),\label{squeeze2}
\end{equation}
where $\mu$ and $\nu$ are functions of squeezing parameter $\xi$
given by: $\mu  = \cosh{|\xi|}$ and
$\nu  = e^{i \phi_{\xi} } \sinh{|\xi|}$.
Using these relations we find that for $\nu/\mu=1$ and $\phi_{\xi}=0$:
\begin{equation}
\mathrm{\hat{D}}(\alpha)
e^{\frac{\hat{a}^{\dagger}\hat{a}^{\dagger}}{2}}
| 0\rangle=\sqrt{\cosh\Gamma}\;|\alpha,-\Gamma\rangle,
\end{equation}
where $\xi=\Gamma=\arctan$h$(1)=\infty$. This means, that left-hand-side of the
equation above corresponds to the infinitely squeezed state. Hence, the
definition of the K--R function given by Eq. (\ref{rih:0}) takes the form
\begin{equation}
K(\alpha,1)=\frac{\sqrt{2}\cosh\Gamma}{\pi}\,
 \mathrm{Tr}\left[\rho |\alpha,-\Gamma\rangle
\langle\Gamma,\alpha|\right].
\end{equation}
In the similar way we can represent the real part of K--R distribution as
\begin{eqnarray}
\mathrm{Re}[K(\alpha,1)]=\frac{\sqrt{2}\cosh\Gamma}{2\pi}\,
 &\mathrm{Tr}& \left[\rho (|\alpha,-\Gamma\rangle
\langle\Gamma,\alpha|  \right.\nonumber\\
&+&\left. |\alpha,\Gamma\rangle
\langle -\Gamma,\alpha|) \right].
\end{eqnarray}
This formula provides a physical interpretation of the K--R distribution
function in terms of a projection of the density operator into a combination
of squeezed states. Let us  consider a linear superposition of
two infinitely squeezed coherent  states
\begin{equation}
|\Phi\rangle={\cal N}^{\frac{1}{2}}
(|\alpha,-\Gamma\rangle+|\alpha,\Gamma\rangle). \label{sq}
\end{equation}
One can easily find that the real part of the K--R distribution, up to a
normalization factor, corresponds to a projection of the density operator on
the off-diagonal elements of the density matrix of the linear superposition
given by $|\Phi\rangle$. This connection with the
coherent squeezed states projection points on an important property of the K--R
function, that will be particularly useful in the phase space visualization of
various properties of squeezed states. The relation of the K--R distribution to
a projection on squeezed states, gives an operational meaning of such a phase
space distribution. The K--R phase space function of a given quantum state at
the phase space point $\alpha$ is just a projection of this state on
 the off-diagonal elements of the density matrix of the
superposition of infinitely squeezed coherent states.

\section{Quantum interference in phase space}\label{sec5}
All phase space functions from the Cohen class of distributions, Eq. (\ref{uog:wig}), are bilinear in
$\Psi$. This provides a transparent exhibition of quantum interference.
For a linear superposition  of quantum states
\begin{equation*}
 |\Psi\rangle=|\Psi_{1}\rangle+|\Psi_{2}\rangle ,
\end{equation*}
the corresponding phase space distribution function takes the form
\begin{equation*}
P=P_1+P_2+P_{int},
\end{equation*}
where $P_1$, $P_2$ correspond to distribution functions of states
$|\Psi_{1}\rangle$ and $|\Psi_{2}\rangle$,  respectively,  and   $ P_{int}$
denotes the interference term. We shall study the properties of the
interference term for the superposition of two plane waves with momenta
 $ p_{1} $ and $ p_{2}$ ($\hbar=1$):
\begin{equation}
\Psi \sim \exp(ip_{1}q)+\exp(ip_{2}q )\;. \label{sup}
\end{equation}
Based on these waves, we shall present a simple argument indicating the
differences between the quantum interference patterns exhibited with the help
of the Wigner function and the K--R function.

The location of the interference terms of the Wigner function is well known.
The Wigner function for such a superposition is
\begin{eqnarray}
W(q,p) & \sim & \delta(p-p_{1})+\delta(p-p_{2}) \label{supwig}\\*
        & + & 2\cos \left(q\negmedspace\vartriangle\negthickspace p\right)
        \delta \left(p-\bar{p}\right),\nonumber
\end{eqnarray}
where $\vartriangle\negthickspace p=p_{2}-p_{1}$ and $\bar{p}=\frac{p_{1}+p_{2}}{2}$.
The characteristic interference term
 $\cos \left(q\negmedspace\vartriangle\negthickspace p\right) $ is
isolated in phase space  from the classical points described by the two momenta
$p_{1,2}$, and is located between these two incoherent terms at the mean
momentum $\bar{p}$.

The generalized K--R function for such a superposition is given by
\begin{eqnarray}
& &K(q,p)  \sim \delta(p-p_{1})+\delta(p-p_{2})\\
& +&
\delta\left(p-\bar{p}-\sigma\negthickspace\vartriangle\negthickspace p\right)
e^{i q \vartriangle\! p}+
\delta\left(p-\bar{p}+\sigma\negthickspace\vartriangle\negthickspace p\right)
 e^{-i q \vartriangle\! p }, \nonumber
\end{eqnarray}
which leads to the following expression  for the real part of the
generalized K--R distribution
\begin{eqnarray}
& &{\mathrm{Re}}[K(q,p)]=\delta(p-p_{1})+\delta(p-p_{2})
\label{suprrih}\\
& +&\cos\big(q\negmedspace\vartriangle\negthickspace p\big)
\Big[\delta\left(p-\bar{p}-\sigma\negthickspace\vartriangle\negthickspace p\right)+\delta\left(p-\bar{p}+\sigma\negthickspace\vartriangle\negthickspace p
\right)
\Big] .
\nonumber
\end{eqnarray}
We see that the interference term still oscillates like
 $\cos\left(q\negmedspace\vartriangle\negthickspace p\right) $,
but the locations of these oscillations have moved to the points
different then in $W(q,p)$, Eq. (\ref{supwig}). Analyzing formula
(\ref{suprrih}) we find that parameter $\sigma$ shifts the
interference term along the momentum axis. For $\sigma=0$, which
corresponds to the Wigner function, interference term is located
at the mean momentum $\bar{p}$. For  $\sigma=1$ (case of the K--R
distribution) these oscillations split and shift to the locations
defined by $p=p_1$ and $p=p_2$ in the phase space. Increase of
$\sigma$ beyond unity will move interference term further apart,
so that they not only no longer overlap each other but appear
outside the physical location of the state.

The main difference between the quantum interference in the Wigner
representation and the K--R representation is the location of the
interference terms. This difference follows from the fact that the K--R
function depends locally on the phase space properties of the wave function,
while for the Wigner function this relation is nonlocal. This is why the
oscillations of the Wigner function occur at a position in phase space
which is different from the classical location.

It may appear, that the K--R function provides a less readable
representation of quantum interference, because the oscillating
terms cannot be isolated from the incoherent location. However,
the advantage of the K--R phase space representation is its local
relationship with the position and momentum wave functions.
 The Wigner distribution function can have oscillations at
points where the wave function is vanishing. From the definition
of the K--R function we see that this phase space distribution has
to vanish when the wave functions $\Psi(q)$ and
$\widetilde{\Psi}(p)$  vanish in $q$ and $p$.

\section{Linear superpositions in the K--R representation}\label{sec6}

\subsection{Coherent states and Fock states}
As an example of  the structures that can be obtained from Eq.
(\ref{k2}) we consider generalized K--R
 distribution function
of a coherent state $|\alpha_0\rangle$.
 By substituting $\hat{\rho}=|\alpha_0\rangle\langle \alpha_0| $ into
Eq. (\ref{k2}) after simple calculations we get the following
formula for generalized K--R distribution function of coherent
state
 $|\alpha_0\rangle$:
\begin{eqnarray}
& &K_{\hat{\rho}}(\alpha,\sigma)  =
\frac{2}{\pi\sqrt{1+\sigma\,^2}}\;  \times  \label{i:sigma}\\*
       & & \exp\left(\frac{\sigma(\alpha^{\star}-
        \alpha_0 ^{\star})^{2} }{1+\sigma\,^2}-
        \frac{2|\alpha -\alpha_0|^{2}}{1+\sigma\,^2}
        -\frac{\sigma(\alpha-\alpha_0)^{2} }{1+\sigma\,^2} \right).
       \nonumber
\end{eqnarray}
For $\sigma=0$ it reduces to the well-known formula for the Wigner
distribution function of a coherent state. For other values of
$\sigma$ the generalized K--R distributions given by Eq.
(\ref{i:sigma}) also consist of a Gaussian bell shape centered
around a point ($\mathrm{Re}(\alpha_0)$, $\mathrm{Im}(\alpha_0)$)
but modified by a phase factor
\begin{equation}
e^{i\delta}=\exp\left[\frac{i 4 \sigma}{1+\sigma^2}
\left( \mathrm{Im}(\alpha)-\mathrm{Im}(\alpha_0)\right)\left(\mathrm{Re}(\alpha)-
\mathrm{Re}(\alpha_0)\right)
\right].
\nonumber
\label{factor1}
\end{equation}
The real part of this distribution has an oscillating term
$\cos\delta$, making the quasi-distribution non-positive.
 For $\sigma\neq 0$, this oscillating term corresponds to the plane wave from
Eq. (\ref{prop2}) with the phase $\varphi(q,p)$ proportional to a
factor $\frac{4 \sigma}{1+\sigma^2}$. With increase of the value of
$\sigma$, these oscillations quickly become more and more rapid,
achieve maximum frequency for $\sigma=1$ and then slowly vanish.

Substituting  $\alpha_0=0$ and $\sigma=1$ into the Eq. (\ref{i:sigma})
 we obtain formula
that describes K--R distribution of the vacuum state $|0\rangle$:
\begin{equation}
K_{|0\rangle\langle 0|}(\alpha,1)  =
\frac{\sqrt{2}}{\pi}\;
       \exp\left(-
        |\alpha|^{2}
        -\frac{\alpha^{2}}{2}+\frac{(\alpha^{\star})^{2}}{2} \right).
        \label{0_fock}
\end{equation}
\begin{figure}
\includegraphics[scale=.73]{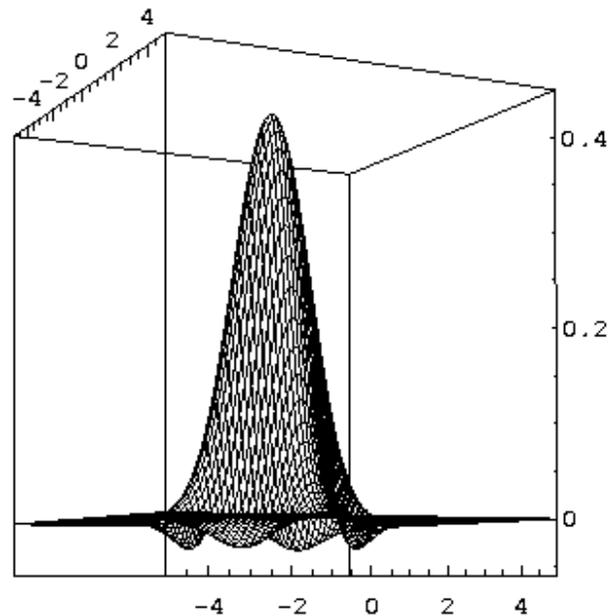}
\caption{ The K--R distribution function for the vacuum state
$|0\rangle$.} \label{fock_zero}
\end{figure}
Figure \ref{fock_zero} shows the real part of the expression given
by Eq. (\ref{0_fock}). In accord with the previous description, it is
a Gaussian function in position and momentum, modulated by the
plane wave phase factor
$\cos[2\mathrm{Re}(\alpha)\mathrm{Im}(\alpha)]$.

\begin{figure}
\includegraphics[scale=.73]{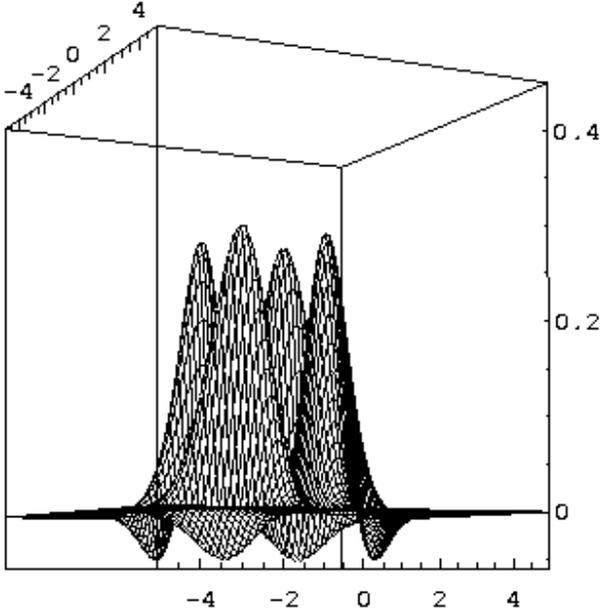}
\caption{ The K--R distribution function for the one photon Fock state. }
\label{fock_one}
\end{figure}

Formula for  K--R distribution of  the one photon Fock state 
$|1\rangle$ is given by
\begin{eqnarray*}
K_{|1\rangle\langle 1|}(\alpha,1)  =
\frac{\sqrt{2}(\alpha^2 - {\alpha^{\star}}^2) }{\pi}\times\\
       \exp\left(-
        |\alpha|^{2}
        -\frac{\alpha^{2}}{2}+\frac{(\alpha^{\star})^{2}}{2}\right).
\end{eqnarray*}
The real part of this equation is shown in the Fig. \ref{fock_one}.
The difference between expressions for the K--R function of the
vacuum state  and the one photon state is an additional multiplicative amplitude $i q  p$ of the oscillating
phase factor. This amplitude comes from the product of the one photon wave
functions in momentum and position representations, respectively.

\subsection{Superposition of coherent states}
Another simple but interesting and instructive example that we shall study is a
superposition of two coherent states $N(|\alpha_0\rangle +
|-\alpha_0\rangle)$. For this state, from Eq. (\ref{k2}) we derive:
\begin{eqnarray}
K(\alpha,\sigma)
=\frac{2 N}{\pi\sqrt{1+\sigma\,^2}}\;
    \left[
    e^{\frac{[\sigma(\alpha^{\star}-\alpha_0^{\star})^{2}-
    \sigma(\alpha-\alpha_0)^{2}-2|\alpha -\alpha_0|^{2}]}
    {1+\sigma\,^2}}
\right.\nonumber\\
+
e^{\frac{[\sigma(\alpha^{\star}+\alpha_0^{\star})^{2}-
    \sigma(\alpha+\alpha_0)^{2}-2|\alpha +\alpha_0|^{2}]}
    {1+\sigma\,^2}}
\nonumber\\
+\left. e^{-2|\alpha_0|^2} \left(e^{\frac{[
    \sigma(\alpha^{\star}+\alpha_0^{\star})^{2}-
    \sigma(\alpha-\alpha_0)^{2}
    -2
    (\alpha^{\star}+\alpha_0^{\star})(\alpha -\alpha_0)]}{1+\sigma\,^2}}
    \right.\right. \nonumber\\
+\left.\left.
e^{\frac{
[
    \sigma(\alpha^{\star}-\alpha_0^{\star})^{2}-
    \sigma(\alpha+\alpha_0)^{2}
    -2
    (\alpha^{\star}-\alpha_0^{\star})(\alpha +\alpha_0)]}{1+\sigma\,^2}}
     \right) \right],
\label{rihkot}
\end{eqnarray}
where
\begin{equation*}
N=\Big(2+2e^{-2\mid\alpha_0\mid^2}\Big)^{-\frac{1}{2}}.
\end{equation*}

\begin{figure}
\noindent
a)
\includegraphics[scale=.7]{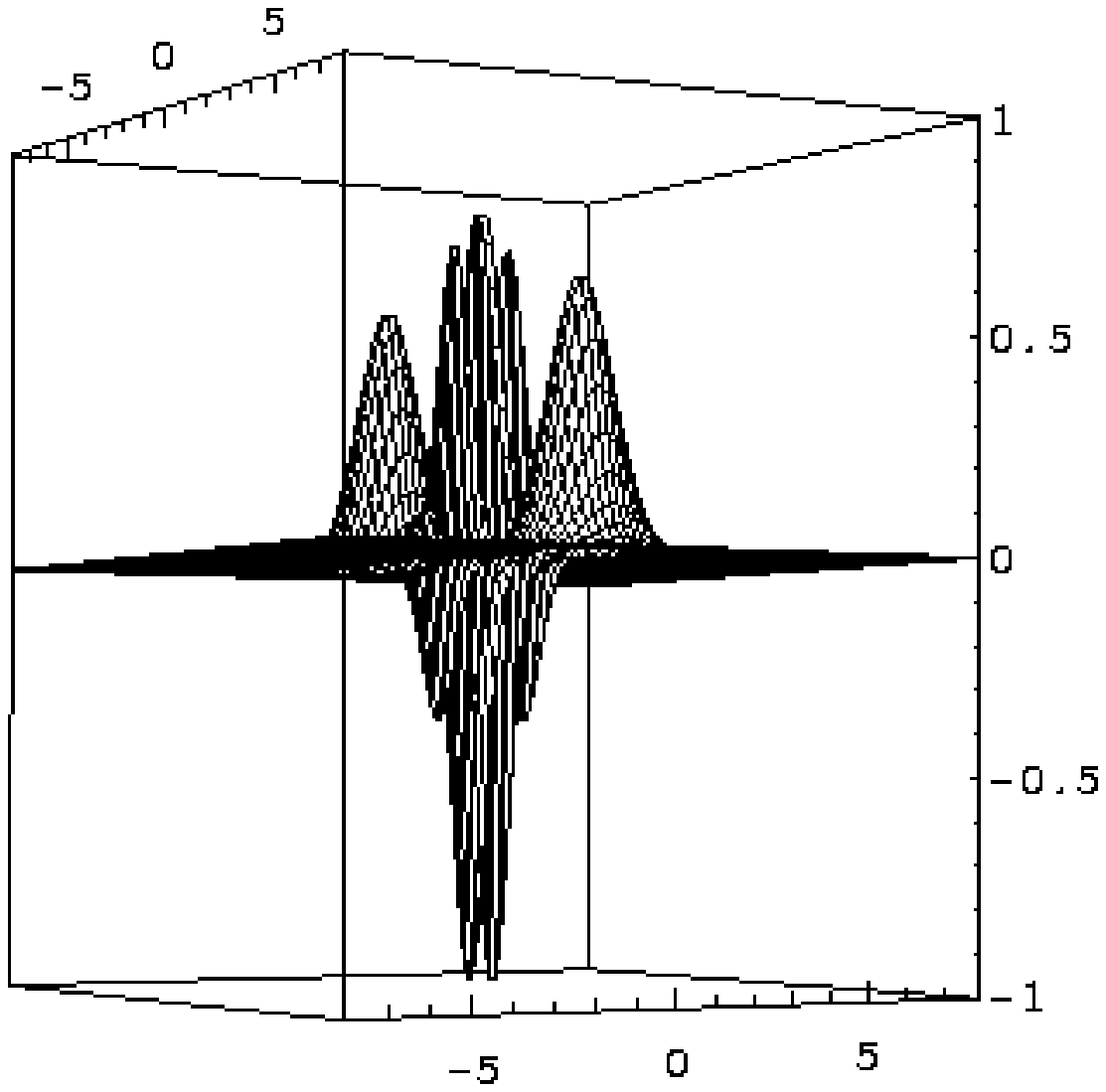}\\\noindent
b)\includegraphics[scale=.7]{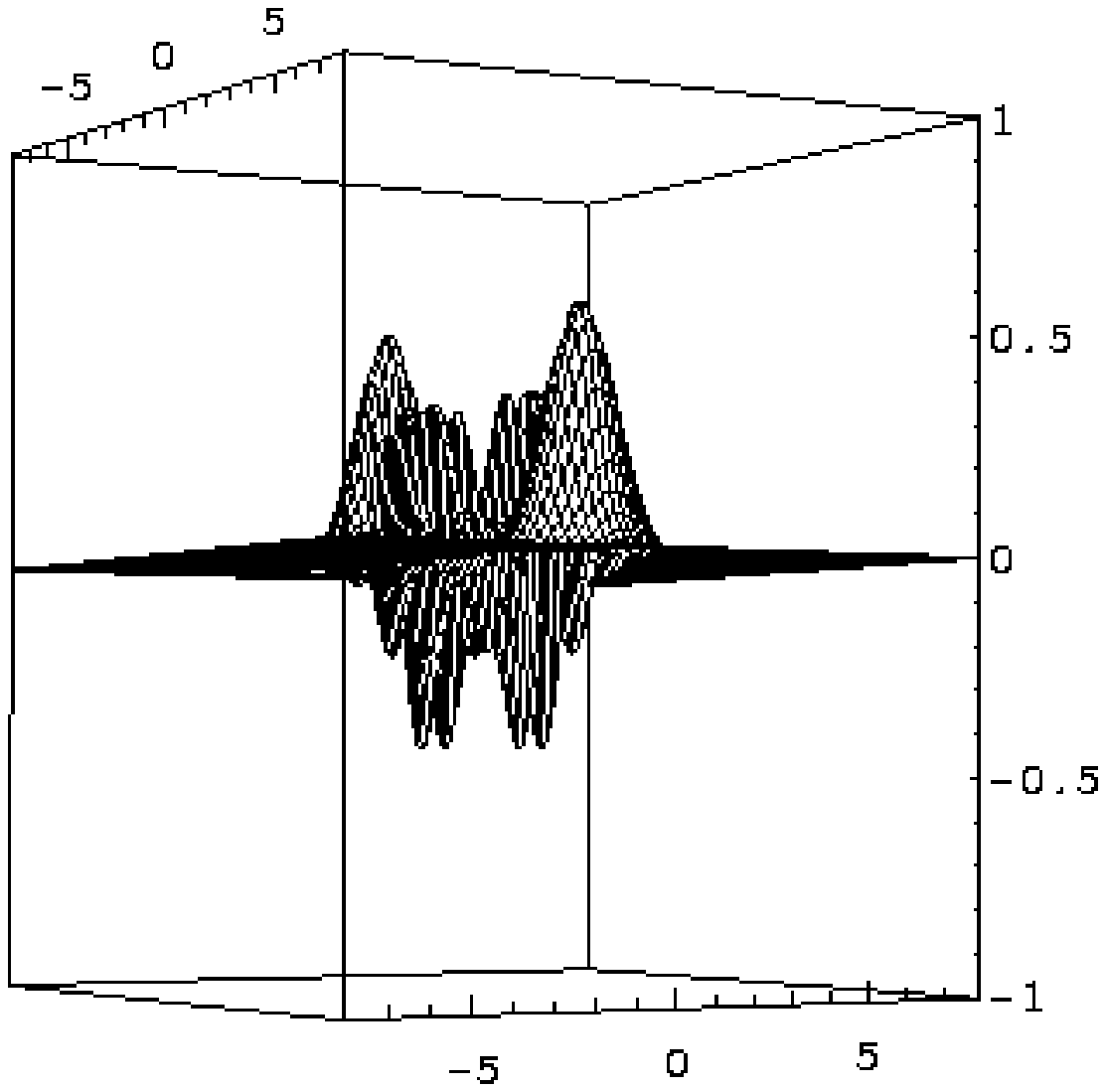} \\\noindent
c)
\includegraphics[scale=.7]{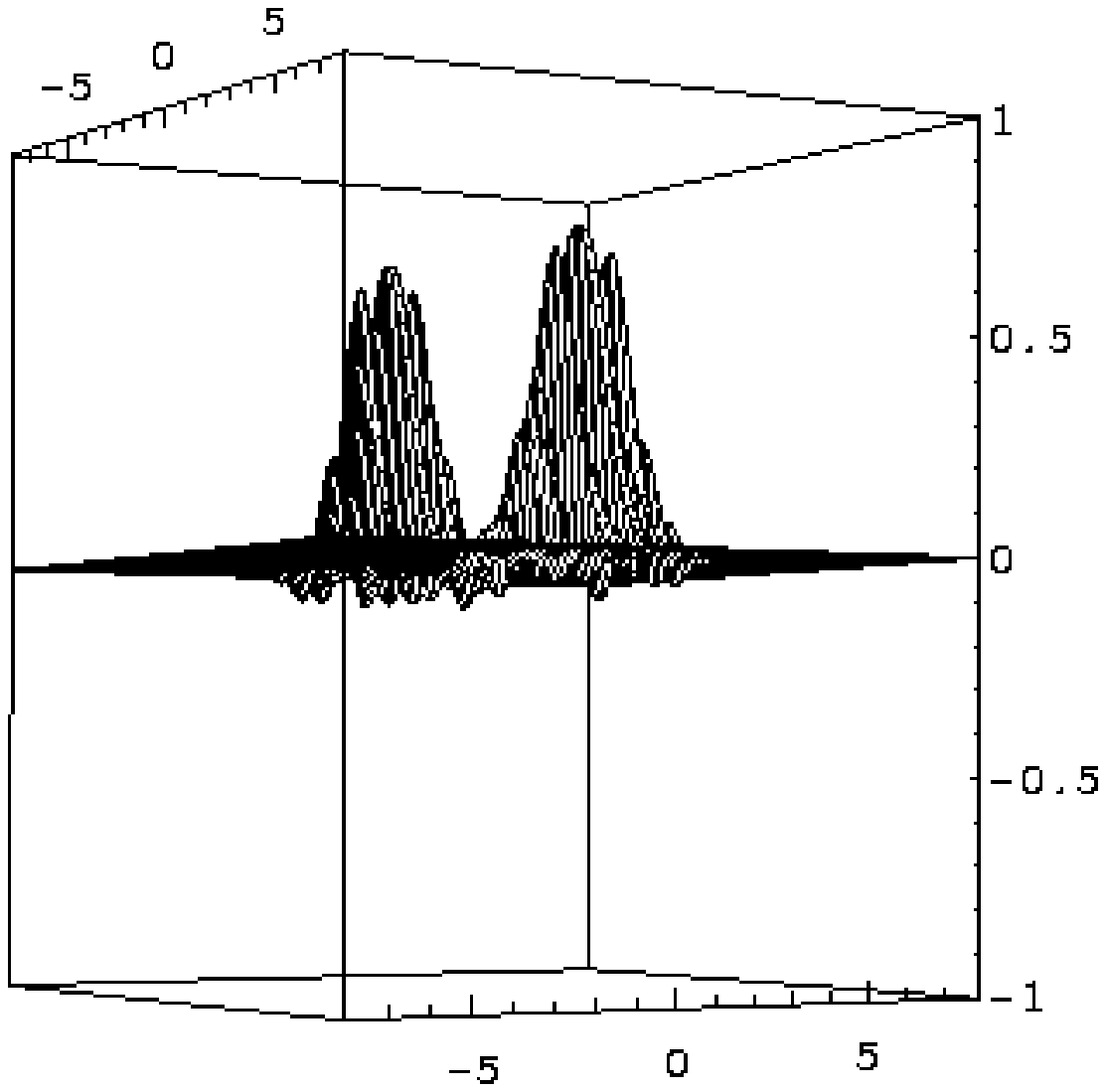} \\
\caption{ Generalized K--R distribution
 function  of superposition of two coherent states
  $N(|\alpha_0\rangle + |-\alpha_0\rangle)$. Here we have chosen
$\alpha_0=3$, and a) $\sigma=0$ (the Wigner distribution function),
b) $\sigma=0.5$, c) $\sigma=1$ (the K--R  distribution function).}
 \label{s0}
\end{figure}

\begin{figure}
$\;\;$\\
$\;\;$\\
a)
\includegraphics[scale=.68]{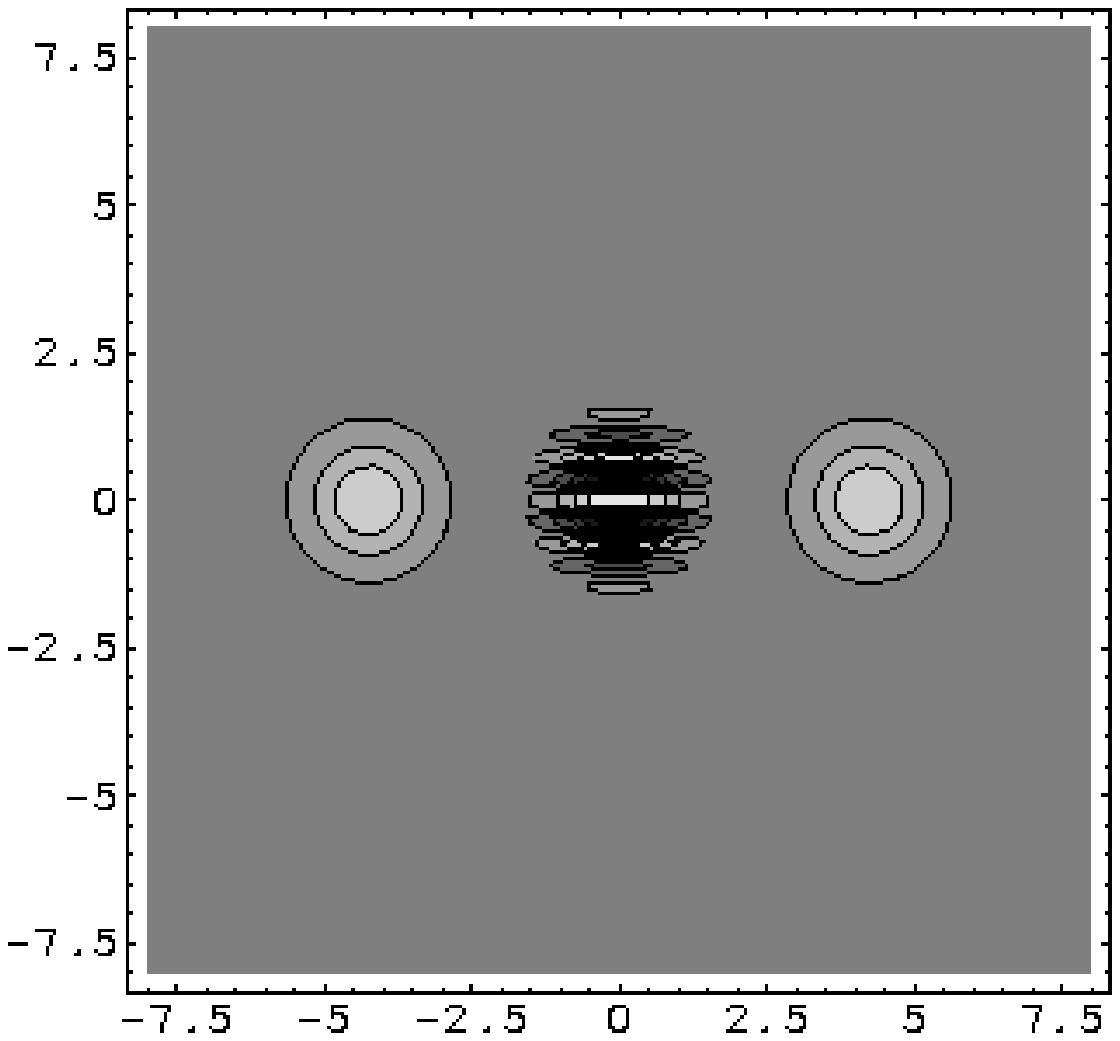}\\
\noindent
b)
\includegraphics[scale=.68]{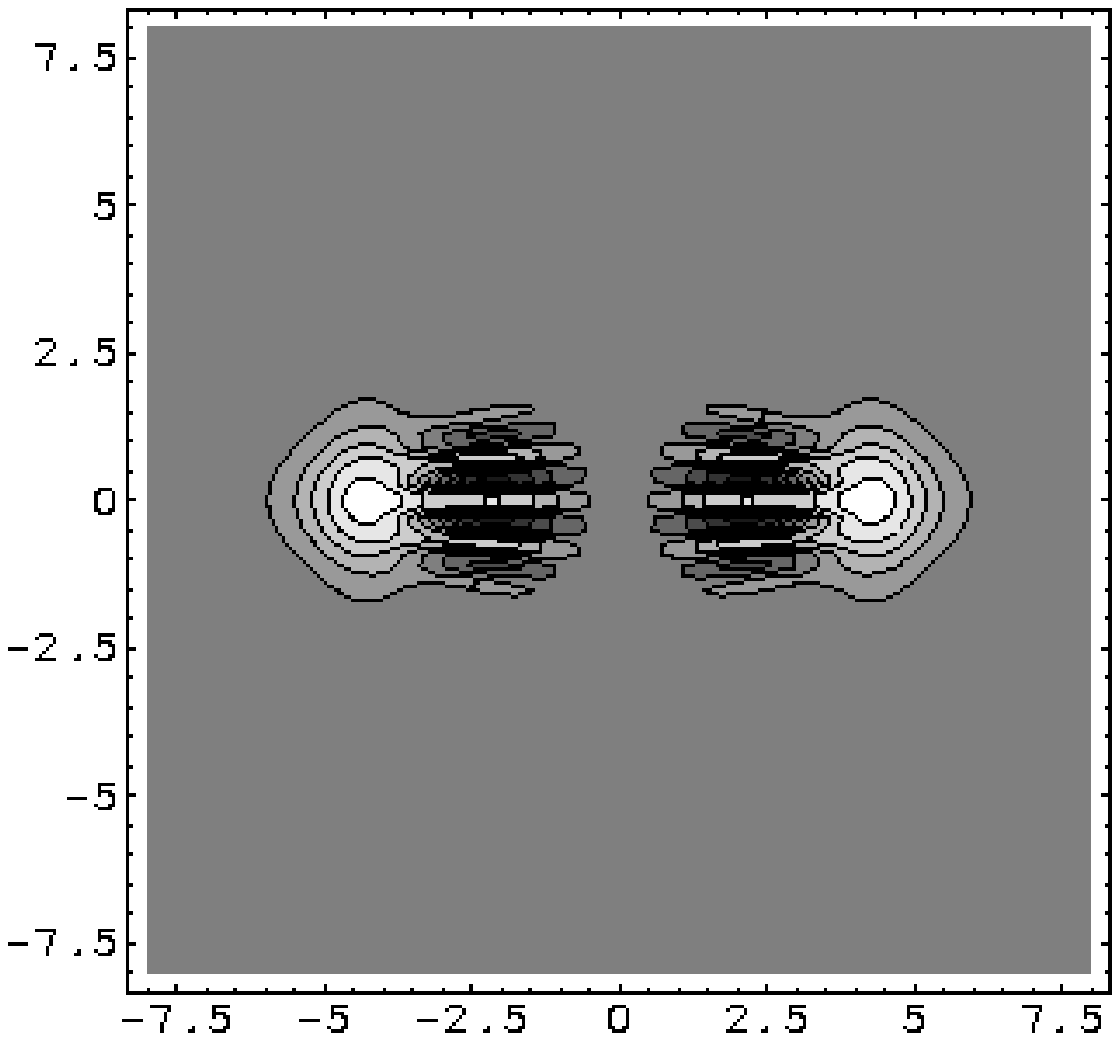}\\
$\;\;$\\
\noindent
c)
\includegraphics[scale=.68]{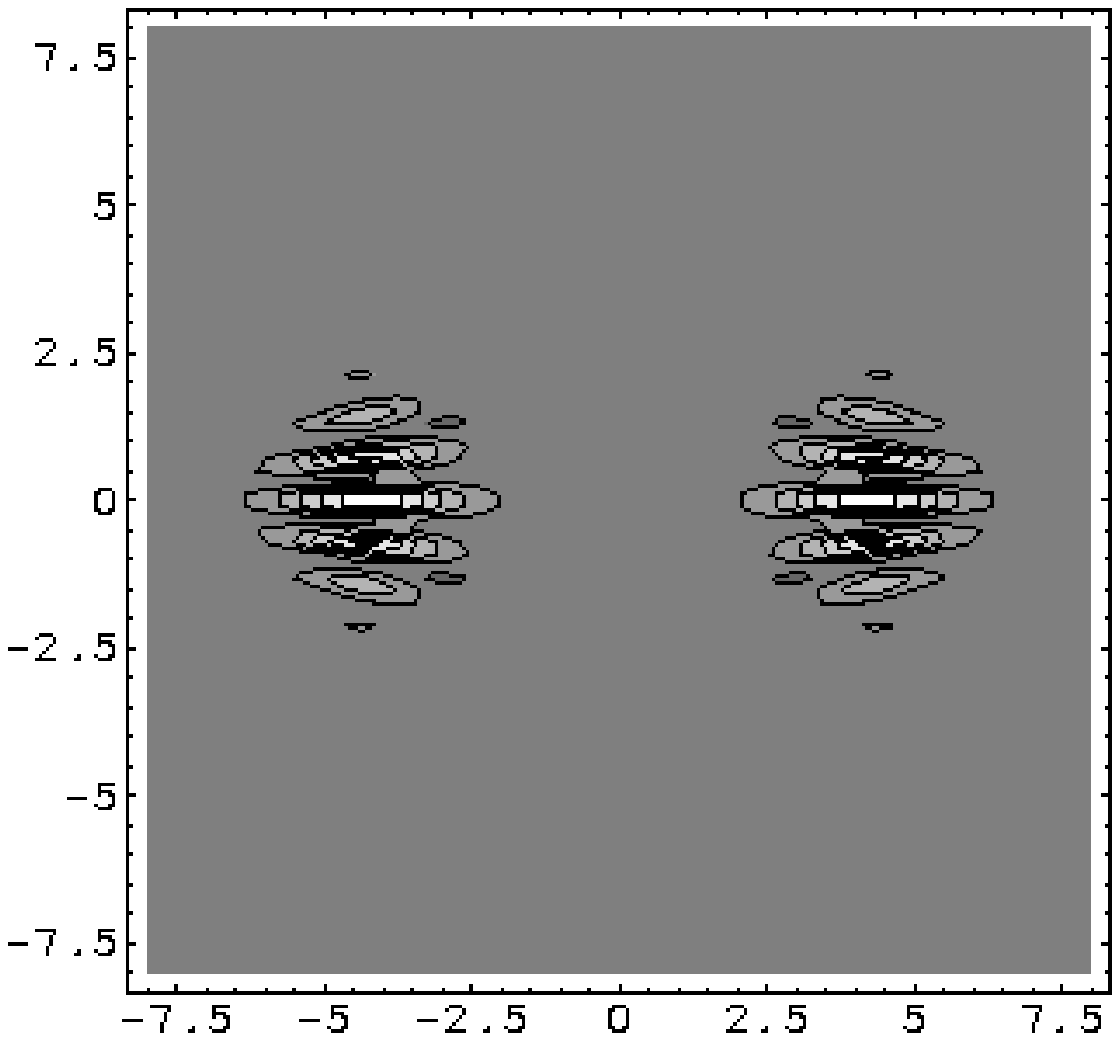} \\
\caption{Contour plots of generalized K--R distribution
function  of superposition of two coherent states
$N(|\alpha_0\rangle + |-\alpha_0\rangle)$.
 Here we have chosen
$\alpha_0=3$, and a) $\sigma=0$ (the Wigner distribution function),
b) $\sigma=0.5$, c) $\sigma=1$ (the K--R  distribution function).}
 \label{s0c}
\end{figure}

Besides terms corresponding to two individual coherent states
there is an interference cross term. Its localization changes with
the change of parameter $\sigma$ in the same way as for the
superposition of two plane waves. This is illustrated in Fig.
\ref{s0}, where we have plotted the real part of Eq.
(\ref{rihkot}) for $\alpha_0=3$ and several values of parameter
$\sigma$. We present  the real part of distributions rather then
the imaginary part (that does not contribute to marginals), or the
absolute value (in which no oscillations appear). In Fig.
\ref{s0c} we present the same functions using contour plots. In
this representation it is apparent how oscillating interference
term moves with the change of the value of the parameter $\sigma$.
As was the case of the superposition of plane waves, for
$\sigma=0$ interference term is located exactly between the terms
that do not correspond to the interference. With the increase of
the value of the parameter $\sigma$   interference terms split
away and move along ${\mathrm{Re}}(\alpha)$ axis, as they are
centered around phase space points $({\mathrm{Re}}(\alpha),
{\mathrm{Im}}(\alpha))=(\pm \sigma\alpha_0,0)$. Again, for
$\sigma=1$ oscillating interference terms overlap the terms
 corresponding to the two individual coherent states.
As we have already mentioned, the  K--R  distribution function is
a special case of a distribution which is always equal to zero when
wave function is zero at certain point.

\begin{figure}
\noindent
a)
\includegraphics[scale=0.68]{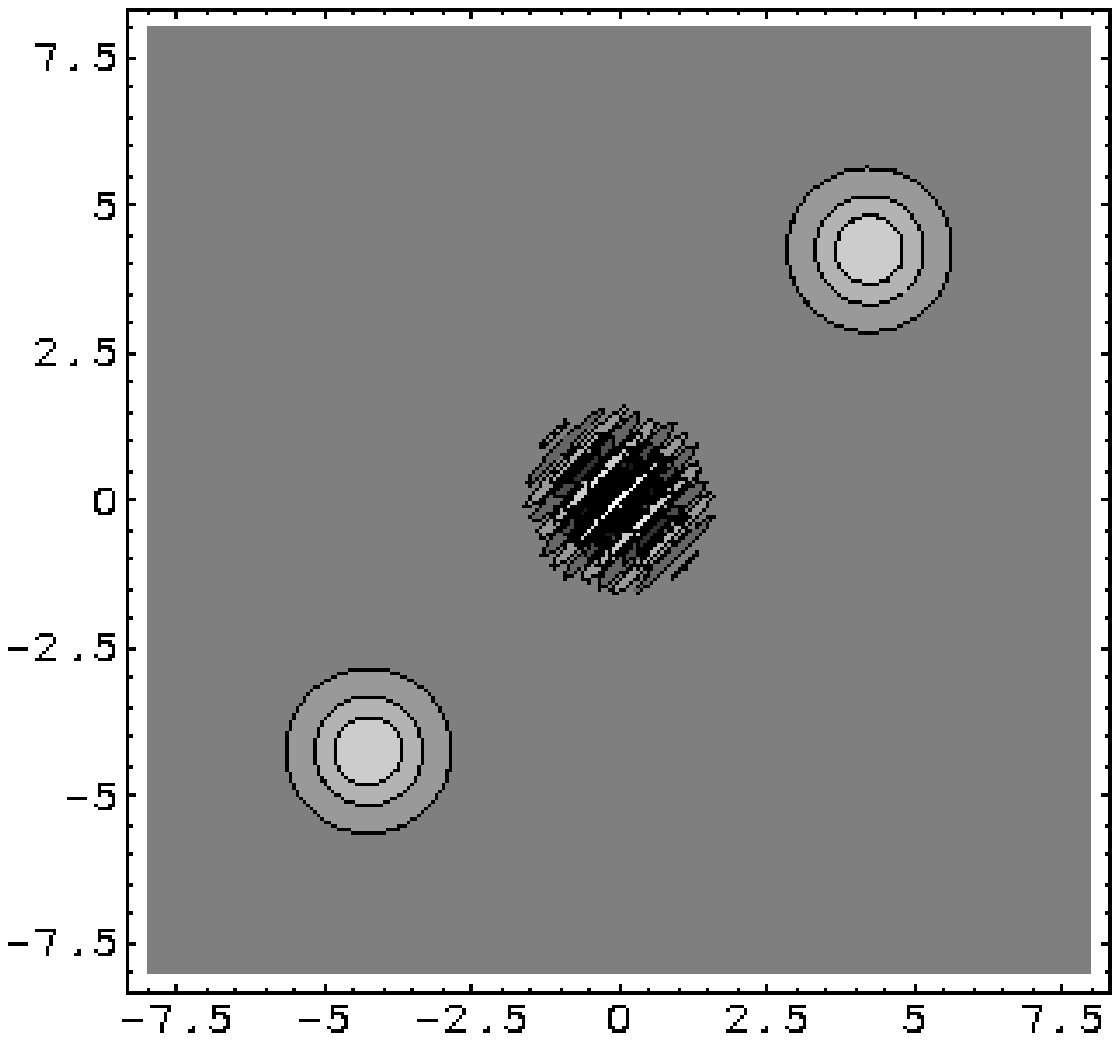}\\
\noindent
b)
\includegraphics[scale=0.68]{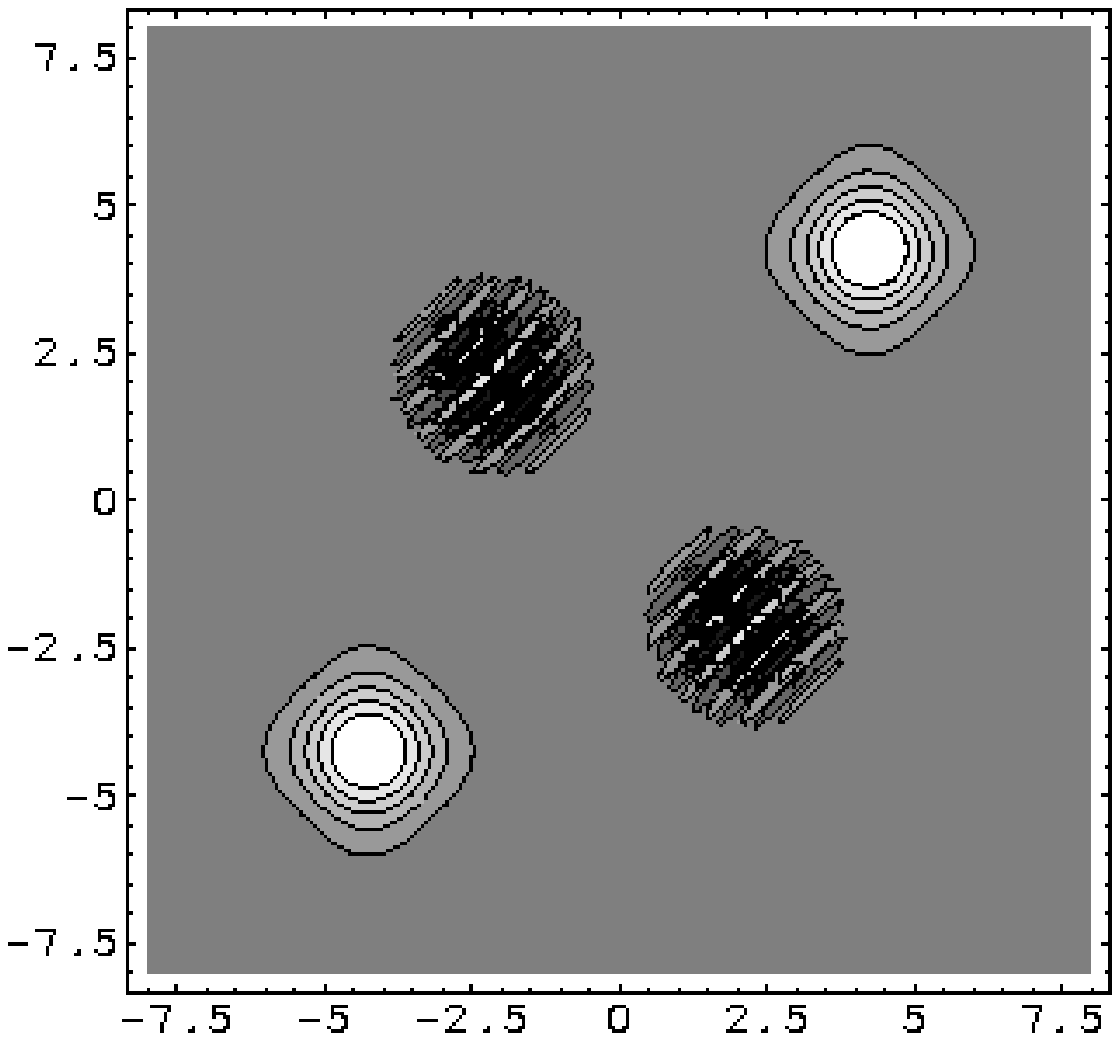}\\\noindent
c)
\includegraphics[scale=0.68]{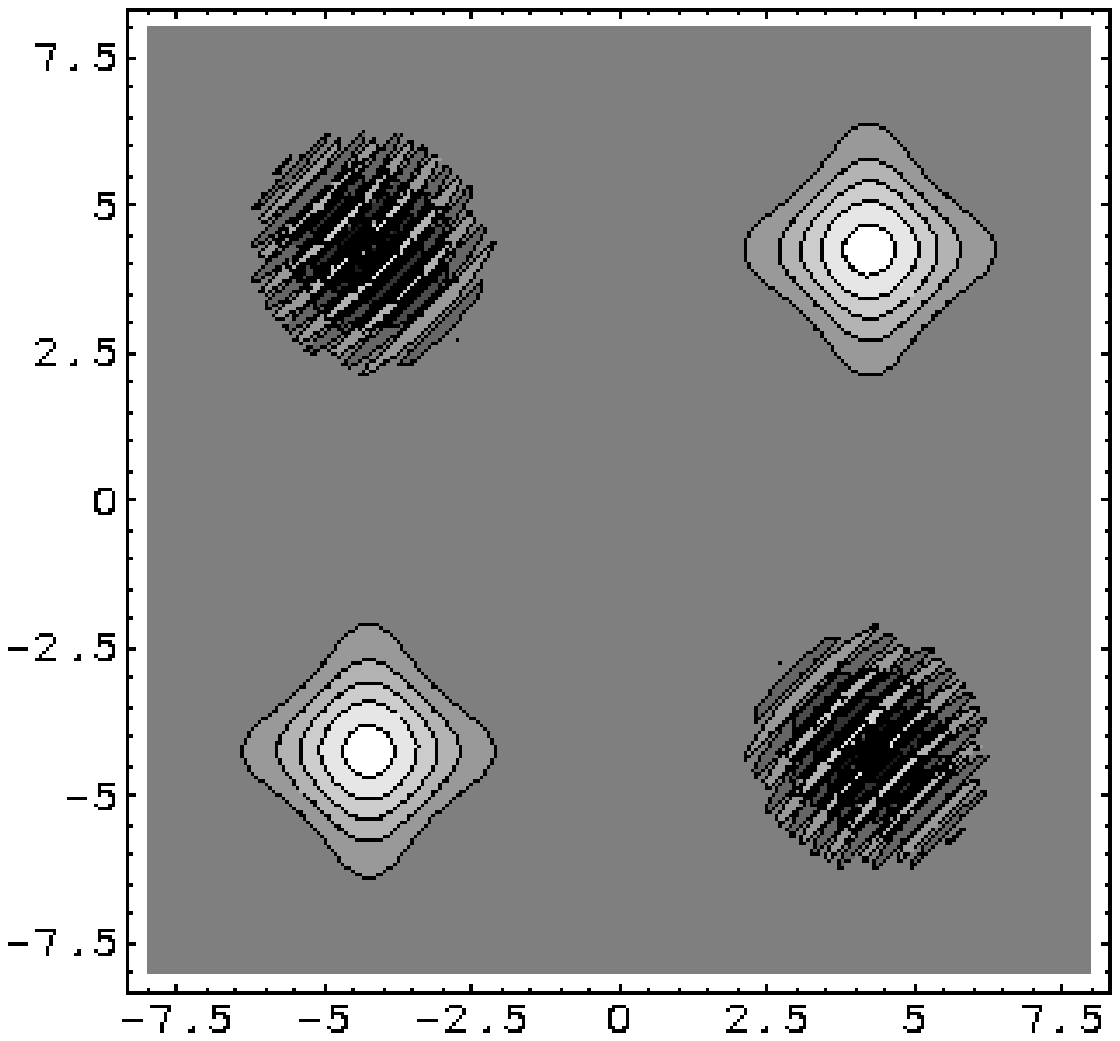}
\caption{Contour plots of generalized K--R distribution
 function  of superposition of two coherent states
  $N(|\alpha_0\rangle + |-\alpha_0\rangle)$. Here we have chosen
$\alpha_0=3+i3$, and a) $\sigma=0$ (the Wigner distribution function),
b) $\sigma=0.5$, c) $\sigma=1$ (the K--R  distribution function). }
\label{s0o}
\end{figure}

Figure \ref{s0o} shows the same distribution functions when the
initial state is  rotated in phase space. As an example we have
chosen previous initial state rotated by $\varphi=\frac{\pi}{4}$
in phase space, i.e. $\alpha_0=3\; e^{i\frac{\pi}{4}}$.
 The Wigner function of
a ``rotated state'' is simply the rotated Wigner function. This
property is the basis for tomography of the Wigner function. By contrast, for other distributions (with parameter
$\sigma\neq 0$) this simple relation does not hold, and tomography
would not work.  

Let us  consider
behavior of the terms corresponding to the individual coherent states in Fig. \ref{s0o}. Although the center of such a term is located exactly in this point
of phase space where coherent state is centered, closer
examination tell us that whole term is not ``rotated'' but rather
``shifted'' by appropriate values parallel to the coordinate axes.
It is clearly seen that for $\sigma\neq 0$ generalized K--R
distribution functions single out $p$ and $q$ axes as compared to
the other directions in phase space. Interference terms are just
rotated but in the opposite direction.

\begin{figure}
$\,$\\
$\xi=0.1$ \\
\includegraphics[scale=0.43]{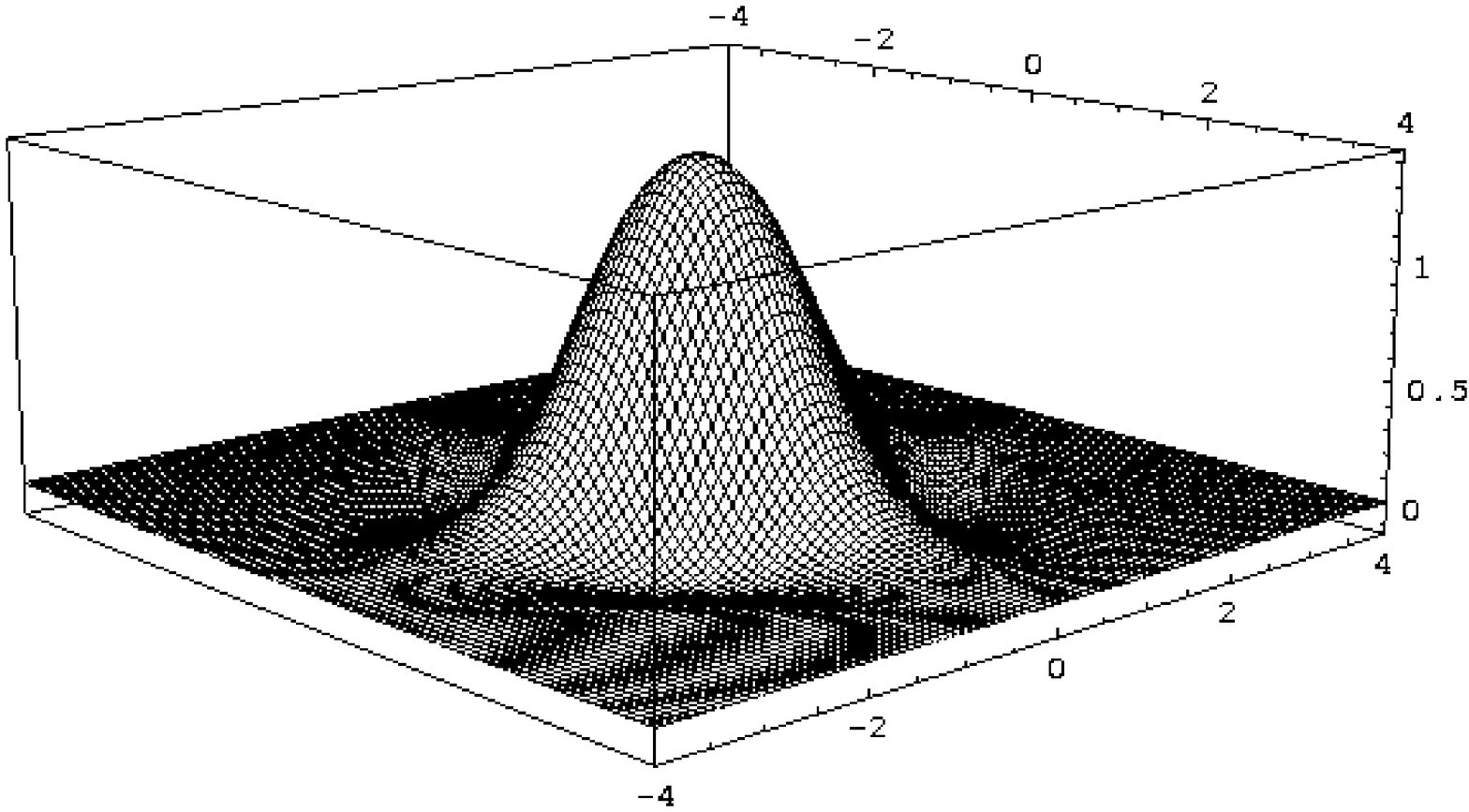}\\
$\,$\\
$\xi=0.5$\\
\includegraphics[scale=0.42]{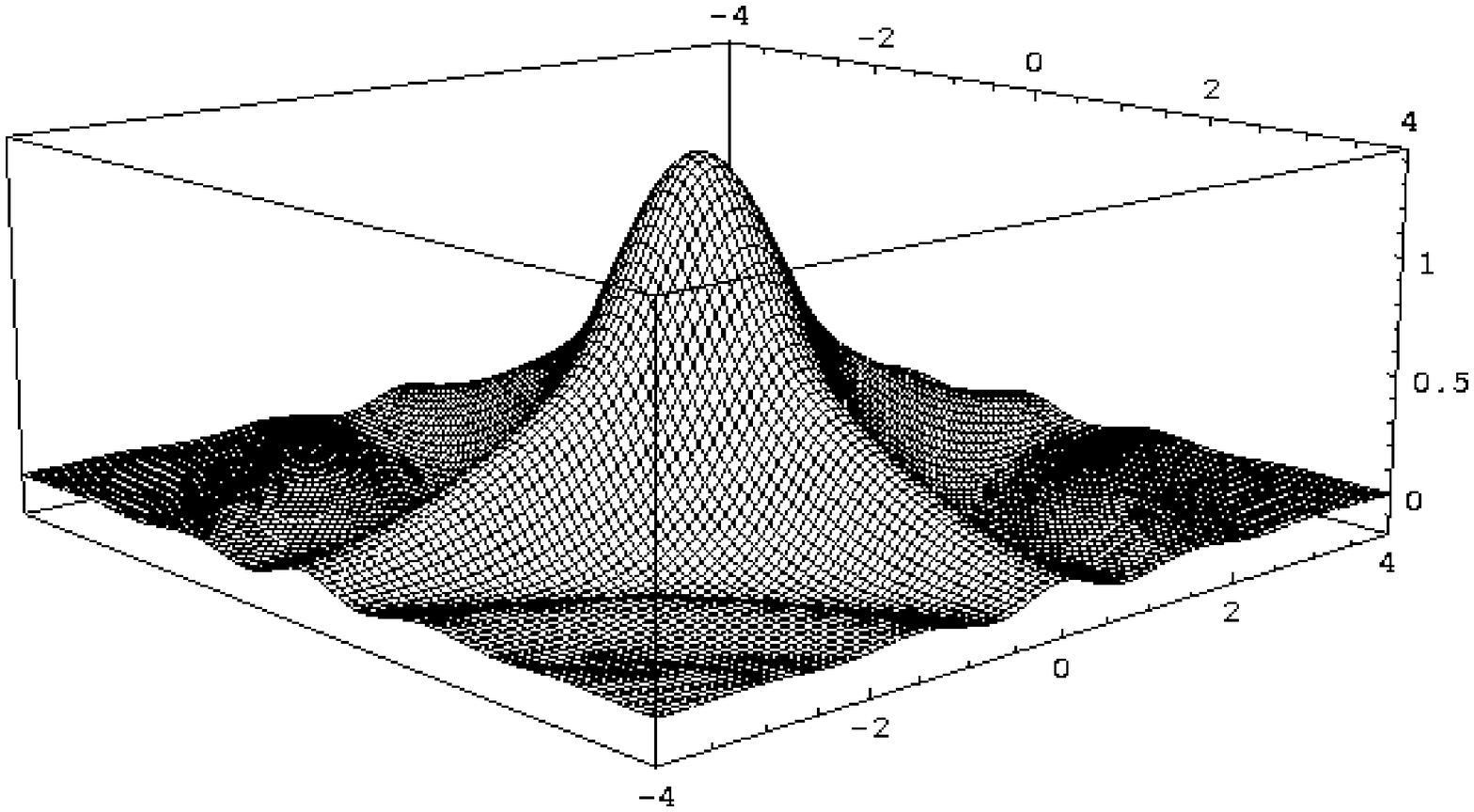}\\
$\,$\\
$\xi=0.7$\\
\includegraphics[scale=0.43]{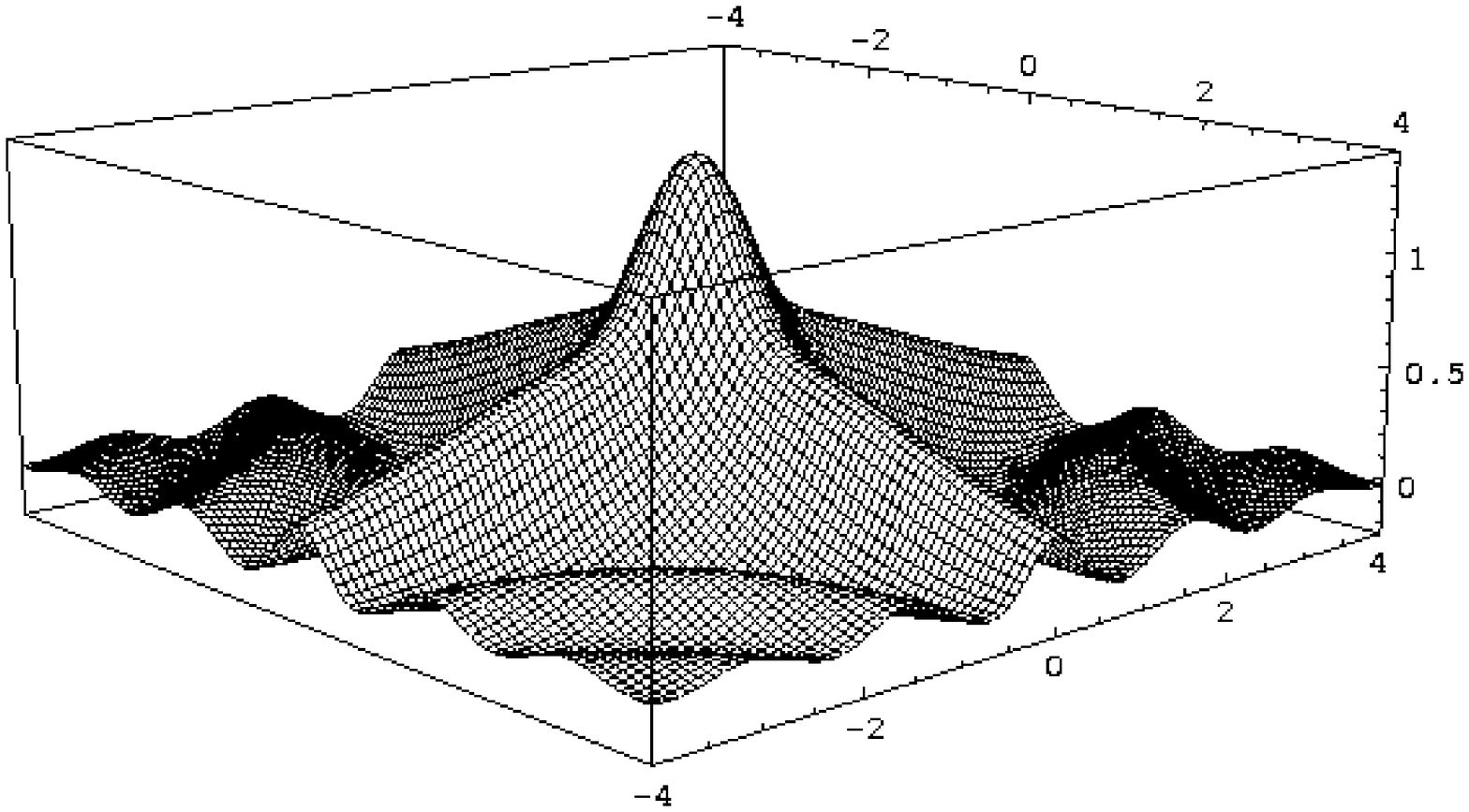}\\
$\,$\\
$\xi=0.9$\\
\includegraphics[scale=0.43]{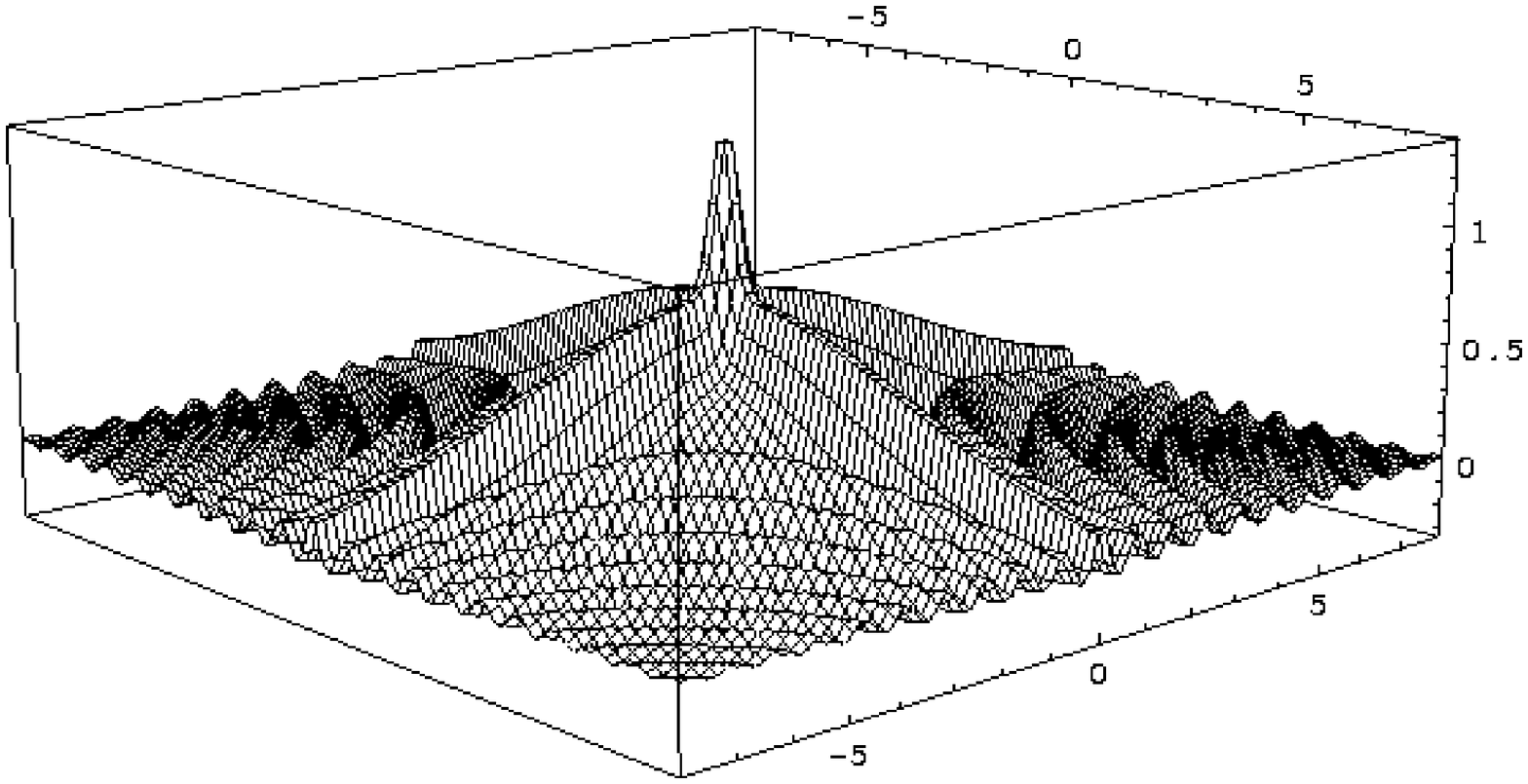}\\
$\,$\\$\,$\\
\caption{
The K--R distribution function of
 superposition of two squeezed coherent states
${\cal N}(|\xi\rangle + |-\xi\rangle)$. }
\label{sqz}
\end{figure}
\subsection{Superposition of coherent squeezed states}

In the previous section we have connected the real part of the K--R function 
with
the expectation value of off-diagonal elements of density matrix
corresponding to
superposition of two infinitely squeezed coherent states. It is interesting to
see how K--R  distribution evolves if we change the squeezing parameter.

In Fig. \ref{sqz} we have plotted the real part of the K--R  distribution function of
superposition of squeezed vacuum states
 ${\cal N}(|\xi\rangle +
|-\xi\rangle)$ for several values of squeezing parameter $\xi$.
The K--R distribution apparently shows that an investigated state
consist of two perpendicularly squeezed states. Obviously, the
closer parameter $\xi$ gets to unity, the more significant are
effects of squeezing. Moreover, also basic properties of K--R
distribution are emphasized with the increase of the value of the 
parameter $\xi$:  the
oscillating $\cos[2\mathrm{Re}(\alpha)\mathrm{Im}(\alpha)]$-like
structure is more and more distinct. As we have mentioned before,
the K--R distribution single out $p$ a $q$ axes from other phase
space directions. Looking at the changes caused by the increase of
squeezing, one gets easily convinced of a close connection between the
K--R distribution and a superposition of infinitely squeezed
coherent states.

\section{Summary}
We have presented a new class of phase space quasi-distribution
functions with correct momentum and position marginal properties,
that contains the Wigner distribution function and the K--R
distribution as special cases. We have shown how quantum
interference appears in phase space if such functions are used for
its investigation. In particular, we have focused on similarities
and differences between the Wigner function and the K--R
distribution.

We have emphasized the most important properties of the K--R
distribution function: the fact that this function fully
characterizes the quantum state; that the K--R function
corresponds to the anti-standard ordering of $\hat{q}$ and
$\hat{p}$ operators; and that the real part of K--R distribution
is in natural way connected to coherent squeezed states.

Most of the properties of the generalized K--R distribution
function we have presented using  the formalism of coherent states
with the full analogy to the  Glauber and Cahill  s-ordered
quasi-distributions.

\section*{Acknowledgments}

This work was partially supported by a KBN grant, {\em Splatanie 
i interferencja atom\'ow i foton\'ow}, and the European Commission
through the Research Training Network QUEST.


\begin{thebibliography}{99}

\bibitem{Wigner}
E. Wigner,
Phys. Rev. {\bf 40}, 749 (1932)

\bibitem{GlaCah}
K. E. Cahill and R.J. Glauber,
Phys. Rev. {\bf 177}, 1857, 1882
(1969)

\bibitem{Gla}
R. J. Glauber,
Phys. Rev. Lett. {\bf 10}, 84 (1963)

\bibitem{Sud}
E. C. G Sudarshan,
Phys. Rev. Lett. {\bf 10}, 277 (1963)

\bibitem{Husimi}
K. Husimi, Proc. Phys. Math. Soc. Japan, {\bf 22}, 246 (1940)\\
K. Takahashi, Suppl. Prog. Theor. Phys. {\bf 98}, 109 (1989)


\bibitem{Kirkwood}
J. G. Kirkwood,
Phys. Rev. {\bf 44}, 31 (1933)

\bibitem{Rihaczek}
A. N. Rihaczek,
 IEEE Trans. Inf. Theory {\bf 14}, 369 (1968)

\bibitem{Englert}
B-G Englert, J. Phys. A, {\bf 22}, 625 (1989)

\bibitem{Zak}
J. Zak,
 Phys. Rev. A {\bf 45},
3540 (1992)

\bibitem{O'Connell}
R. F. O'Connell and E. P. Wigner, Phys. Lett. A vol 83 no 4 (1981), 145

\bibitem{ff}
W. P. Schleich, {\em Quantum Optics in Phase Space}, WILEY-VCH (2001).


\bibitem{Cohen0}
L. Cohen, Proceedings of the IEEE {\bf 77}, No. 7 (1989).

\bibitem{Coh01}
L. Cohen, {\em Time-Frequency Analysis}, PRENTICE  HALL (1995).


\bibitem{M-H}
H. Margenau, R. N. Hill, Prog. Theor. Phys., {\bf  26}, 722 (1961)


\bibitem{new}
L. Praxmeyer, K. W\'odkiewicz, {\sl Hydrogen atom in the Kirkwood--Rihaczek
representation} (in preparation)


\bibitem{Cohen}
L. Cohen,
J. Math. Phys. {\bf 7}, 781 (1966)


\bibitem{Cohen2}
L. Cohen and Y.I. Zaparovanny,
J. Math. Phys. {\bf 21}, 794 (1990)



\bibitem{AgW}
G. S. Agarwal and E. Wolf, Phys. Rev. D {\bf 2}, 2161, 2187, 2206
(1970)

\bibitem{squeeze}
D. F. Walles, G. J. Milburn, {\em Quantum Optics}, SPRINGER--VERLAG
(1994)

\end{thebibliography}
\end{document}